\documentclass[3p,times,twocolumn]{elsarticle}

\usepackage{ecrc}
\usepackage{hyperref}

\volume{00}

\firstpage{1}

\journalname{Nuclear Physics B Proceedings Supplement}

\runauth{Diego Gallego}

\jid{nuphbp}

\jnltitlelogo{Nuclear Physics B Proceedings Supplement}

\usepackage{amssymb}

\usepackage[figuresright]{rotating}
\usepackage{float}

\begin{document}

\begin{frontmatter}



\dochead{}

\title{Effective upliftings in Large volume compactifications}


\author{Diego Gallego}

\address{Escuela de F\'isica, Universidad Pedag\'ogica y Tecnol\'ogica de Colombia (UPTC)\\Avenida Central del Norte, Tunja, Colombia}
\ead{diego.gallego@uptc.edu.co}
\begin{abstract}
After reviewing several mechanisms proposed to get a dS/Minkowski
vacuum in moduli stabilization scenarios of type-IIB superstring
orientifold compactifications we propose a criterium for
characterizing those that may effectively lead to a positive small
cosmological constant. We suggest that the variation in the
expectation value of a good uplifting term, due to the shift in the
minimum of the potential after uplifting, is much smaller than the
original cosmological constant. This is studied with some detail in
Large volume scenarios where the dependency on the volume direction
is rather generic and easy to spot. Here we find that an uplifting
term in the potential, with generic form $V_{up}\sim 1/{\cal
V}^\gamma$, should be restricted to the one satisfying
$\gamma^2\ll12$. Such a bound might explain why in models previously
studied no uplifting has been achieved, and gives motivations to
study a novel proposal of dilaton dependent uplifting mechanism for
which no numerical studies has been performed before. We find that
in this case it is actually possible to get a dS vacuum, but still
leave open the question of a more precise discrimination feature the
good uplifting mechanisms should satisfy.
\end{abstract}

\begin{keyword}
String compactifications, dS vacua in  string theory.


\end{keyword}

\end{frontmatter}


\section{Introduction}
In the LHC era superstring theory seems to be in good shape to
accept the challenge: great deal of progress in recent years has
shown that it is possible to reproduce the particle spectra of the
Standard Model o well motivated
extensions\cite{MINILand1,MINILand2,MINILand3,DModels1,DModels2}, as
well scenarios with enough dynamics to stabilize all moduli with low
energy Supersymmetry \cite{KKLT} leaving room for cosmological
features like inflation \cite{SInfRev}. In this context the so
called Large volume scenarios (LVS) \cite{LVS} present a kind of
type-IIB vacua where both features has been shown to be present
\cite{LVSLocMod1,LVSLocMod2,LVSLocMod3}, thus, despite of being more
rigid in implementation compared to models inspired by the seminal
work of Kachru, Kallosh, Linde and Trivedi in \cite{KKLT}, these a
unique instance worth to be developed.
\\
One of the features that moduli stabilization should give, if it is
to help with the cosmological hierarchy problem, is the realization
of a nearly vanishing positive value for the cosmological constant
\cite{CCExp1,CCExp2}. Although the natural scenario for superstring
vacua is a deep AdS, toy models have been proposed obtaining
dS/Minkowski minima without the need of anti-D branes in the
description
\cite{Ftermup0,Ftermup1,Ftermup2,Ftermup3,Ftermup4,Ftermup5,Ftermup6,Ftermup7,Ftermup8,Ftermup9,Ftermup10,Ftermup11,Ftermup12,Ftermup13,Dtermup1,Dtermup2,Dtermup3,Dtermup4,Dtermup5,Kahlerup1,Kahlerup2,Kahlerup3,Kahlerup4,LVSMatter,LVSMatterInvolved}.
For the LVS only a couple of works have done the honest work of a
numerical computation of such vacua
\cite{LVSMatterInvolved,mioLVS,Ftermup12}, while the rest remain
relaying in parametrical indications which in any case do not
guarantee the final implementation. Then, for instance, in a model
proposed by Cremades et al. in \cite{LVSMatter}, an F-term from a
matter-like field serves as uplifting for the cosmological constant.
However, later studies by the author showed that such model fails in
getting an dS/Minkowski \cite{mioLVS}. In other words the uplifting
procedure does not achieve its target.
\\
The intention of this letter is to give a first proposal for a
characterization that identifies good uplifting mechanisms in the
LVS. For this we concentrate on the uplifting term in the potential,
which we define as one with the property of being positive definite
and depending on parameters whose tuning allows, in principle, for a
cancelation of the originally negative cosmological constant. We
propose that an effective uplifting mechanism is one for which the
difference between the Vacuum Expectation Value (VEV) of the
uplifting term at the unperturbed minimum, when the uplifting is
turned off, and its VEV in the perturbed one, when the uplifting is
turned on, is much more smaller than the VEV of the potential
without uplifting. In order to be more quantitative we use the fact
that in general the uplifting part of the scalar potential comes as
a negative power of the overall volumen, i.e., $V_{up}\sim {\cal
V}^{-\gamma}$, for which we find, implementing the condition above,
a bound given by $\gamma\ll 12$. Such a bound is latter on studied
numerically. This results not only helps in understanding the
failure of previous models but also as guidance for future
constructions.
\\
The letter is organized as follows: the next section reviews some
standard mechanisms for generating an uplift. This is used in
section \ref{proposalsec} to presents and justify the requirement
proposed in the paper, as well the explicit implementation in the
case of the LVS. The last part of the letter discuses two uplifting
scenarios, one for which a dS/Minkowski is not realized, something
that is understood in the light of the requirement just presented.
The last section serves as discussion on the results, which,
however, are only a small part of broader and more systematic study
left for a future report.

\section{Uplifting terms}
We work in the framework of ${\cal N}=1$ Supergravity (SUGRA) as the
low energy effective description of type-IIB superstring theory
compactified on a Calabi-Yau 3-fold with an Orientifold plane. Then
the different sources of uplifting come either from the F or D-term
part of the scalar potential, or from stringy, Supersymmetry (SUSY)
breaking, effects like anti-D branes.

\subsection{Anti-brane potential}
In some situations, in order to satisfy the tadpole condition which
depend on the fluxes and the net D-brane number, a number of anti-D
branes are included in the setup. In the probe approximation it is
possible to get an expression for their contribution to the energy
\cite{antiDbrane}, proportional to the brane tension
\begin{equation}
V_{\overline{D}}=\frac{\nu}{{\cal V}^\gamma}\,,
\end{equation}
with the parameter $\gamma$ depending on the type of brane in the
game and the amplitude $\nu$ going like a warping factor evaluated
at the end of the throat, where the anti-brane is localized.
\\
This uplifting mechanism is quite robust being completely
independent of the scalar potential stabilizing the moduli, thus has
been the main way used for justifying dS/Minkowski vacua. However,
since it breaks explicitly SUSY its low energy SUGRA implementation
seems sometimes obscure (see however \cite{Kallosh:2014wsa}).

\subsection{F-term uplifting}

A more controlled way of introducing positive terms is adding new
degrees of freedom that contribute to the F-term scalar potential,
\begin{equation}\label{FtermPot}
V_F=e^{K}\left(\overline{D}_{\bar J}\overline W K^{\bar J
I}D_IW-3|W|^2\right)\,,
\end{equation}
in $M_{Planck}=1$ units, with $W$ and $K$ the superpotential and
K\"ahler potential respectively, $D_IW=\partial_IW+W\partial_IK$ the
covariant derivative, the indices $I$ and $J$ running over the
fields $\phi_I$, and $K^{\bar J I}$ the inverse of the scalar
manifold metric $K_{I\bar J}$. Indeed the diagonal terms, $K^{I\bar
I}|D_IW|^2$, being positive definite might potentially uplift the
vacuum.
\\
An ubiquitous sector in string compactifications are matter like
fields, for which the K\"ahler potential behaves like
\cite{KahlerMatter}
\begin{equation}
K_{matter}\sim {\cal V}^{-\eta}|\phi|^2\,.
\end{equation}
Then regarding the moduli K\"ahler potential as $K_{mod}\supset- 2
ln{\cal V}$, a term of the form
\begin{equation}
V_{matter}\supset \frac{1}{{\cal V}^{2-\eta}}|D_\phi W|^2\,,
\end{equation}
is induced in the scalar potential, which depends on the VEV for the
matter field dictated by other dynamics, usually the D-term
potential. The extra degrees of freedom can be as well other moduli
that can be easily stabilized by other means, like in the model
presented later on in sec.(\ref{Dilatondepsec}). This kind of
uplifting as been exploited in several works so far
\cite{Ftermup0,Ftermup1,Ftermup2,Ftermup3,Ftermup4,Ftermup5,Ftermup6,Ftermup7,Ftermup8,Ftermup9,Ftermup10,Ftermup11,Ftermup12,Ftermup13}.
\\
Another possibility for this kind of uplifting is to consider
perturbative corrections, $\alpha^\prime$ or string loops, to the
K\"ahler potential which might lead to important contributions
potentially able to uplift the vacuum
\cite{Kahlerup1,Kahlerup2,Kahlerup3}. In this case a careful
analysis on higher order corrections is compulsory as these might
invalidate the initial first order implementation.
\\
Contrary to the case of anti-D brane uplifting in here the moduli
stabilization process should be, in principle, done simultaneously
to uplifting, as the extra degrees of freedom or the quantum
corrections might affect completely the way the moduli are
stabilized. This is why some authors prefers not to call this as
uplifting. However, in almost all the cases it is possible to
identify an uplifting term satisfying the definition above.

\subsection{D-term uplifting}

Once the dynamics includes gauge symmetries an extra contribution to
the scalar potential appears. In general one can work in a mesonic
SUSY preserving branch, such that the the only source of SUSY
breaking and uplifting comes from $U(1)_X$ sectors, for which this
contribution reads
\begin{equation}
V_D=\frac{1}{2Re(f_X)}(i\chi^I_X\partial_IK)^2\,,
\end{equation}
with $\chi^I_X$ the Killing vectors of the symmetry and $f_X$ is the
field dependent gauge kinetic function corresponding to the gauge
sector. Matter like fields transform linearly,
$\chi^\phi_X=iq^\phi_X \phi$, proportional to the charge, while the
moduli get charged non-linearly, i.e., $\chi^{\cal
M}_X=i\delta_X^{\cal M}$ with $\delta_X^{\cal M}$ a constant, once
magnetic fluxes are turned on in the D-branes supporting the matter
fields and wrapped in the cycle parameterized by the moduli
\cite{FI1}. Once the VEV for this part of the potential is not
vanishing it leads to a positive definite contribution that might
uplift the vacuum.
\\
Given the dependency of the K\"ahler potential on the volume, as
well possible dependencies in the gauge kinetic function, the
induced uplifting behaves like $V\sim 1/{\cal V}^\gamma$.
\\
Implementing this scenario, although not imposible
\cite{Dtermup1,Dtermup2,Dtermup3,Dtermup4,Dtermup5}, is in general
more involved given dynamical constrains on the the D-term potential
\cite{GomezReino:2007qi}.

\section{The requirement}\label{proposalsec}
\begin{figure}[h!]
 \centering
      \includegraphics[width=0.35\textwidth]{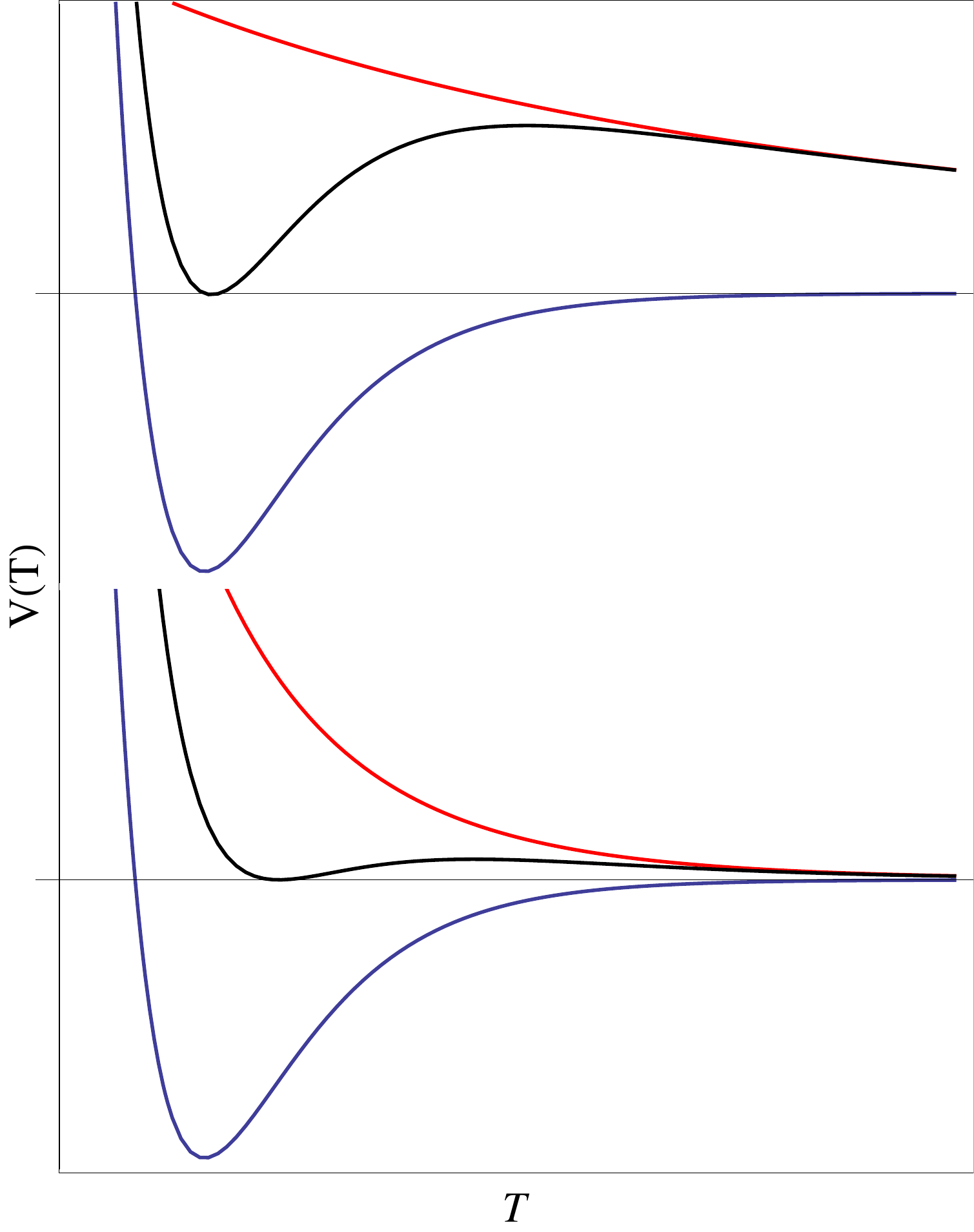}
  \caption{Single modulus case. The original potential (blue) is uplifted by $V_{up}\sim 1/T^n$ (red). Case on top with power $n=2$, and bellow with $n=11$. The total potential (black) is flatter for a steeper uplifting and the shift in the position of the minimum is larger.}
\label{potentialfig}
\end{figure}

The outshot of the previous section is the fact that the sources of
uplifting in general generate uplifting terms that have a runaway
behavior in the compact manifold volume direction, namely its
perturbation on the vacuum goes in the de-compactification
direction. Then, even if the uplifting at the end does not wipe out
the vacuum the resulting potential is flatter (see figure
\ref{potentialfig}), so that the masses for the moduli in general
get lowered compared the ones obtained in the initial stabilization
potentially facing the cosmological moduli problem.
\\
One might still try to construct models with high SUSY breaking
scale and testable signatures encoded in this light modulus
direction, but a large perturbation on the vacuum position might
invalidate the analysis and understanding of the stabilization
process that is what allows the control for possible modifications
in order to generated desirable physics.

Requiring a small perturbation is still not enough for a precise
characterization of a good uplifting since this term should be
relevant enough to do its job, namely, it should be comparable to
the original potential so to cancel the cosmological constant. Let
us illustrate the situation in the context of the numerical
computation of the vacuum position and its properties; in here  a
tuning of some free parameters in the uplifting term allows, in
principle, to get a vanishing cosmological constant. The value for
these then is dictated by the value of the VEV of the original
potential. However, due to the perturbation from the uplifting this
VEV is shifted to larger values of the volume and the values for the
parameters might not be the correct ones as the uplifting term is
smaller the larger the volume. Changing the values for the
parameters translates then in a larger shift in the vacuum that
again might or might not be uplifted. The game then is iteratively
done until, hopefully, a dS/Minkowski vacuum is obtained. In some
cases at the end the perturbation is so large that the potential is
flat enough that spotting the vacuum is quite hard or, in the worst
case, its properties are changed completely.\footnote{For example by
the collapse with neighbor solutions that might not be minima.}
Actually if the uplifting term decreases faster than the original
potential the minimum will be always an AdS no matter how large is
its constant amplitude modulated by the tuned parameters.
\\
A good uplifting, then, avoids such problems by a ``quick'' uplift
which, however, still can be thought as a small contribution in the
stabilization process. This, we propose, is encoded in following
condition
\begin{equation}\label{const}
|\langle V_{up}\rangle_0-\langle V_{up}\rangle|\ll |\langle
V_{0}\rangle|\,,
\end{equation}
where we denote by $V_{0}$ and $V_{up}$ the original potential and
the uplifting term respectively and $\langle \rangle_0$ the VEV
evaluated at the unperturbed vacuum while the other VEV's are
evaluated at the point resulting by introducing the uplifting with
parameters that in principle lead to a cancelation of the
cosmological constant. Condition (\ref{const}) simply requires that
the shrink on the uplifting, due to the shift, is small enough so
that the uplifting indeed proceeds. This together with a precise
knowledge of the stabilization procedure can help in establishing, a
priori, if a possible uplifting will indeed do the job or not.
\\
In general the situation can be complicated enough so that in any
case such an analysis is not straightforward. We simplify the
analysis by assuming the uplifting as a function of the volume
alone. Although this is not exactly the case a justification for
such an assumption is that in LVS the volume modulus is sightly
lighter than the rest of directions, meaning that the dynamics
responsible of stabilizing the other directions are parametrically
larger and, therefore, the direction that is going to be more
affected by the uplifting is precisely the volume one. Having said
that and regarding a small shift we can quantitatively evaluate its
value by expanding the equation of motion for $V=V_0+V_{up}$ around
the original minimum, given by $\partial_{{\cal V}}V_0=0$, leading
to
\begin{equation}\label{shiftGen}
\delta {\cal V}\approx \left\langle\frac{\partial_{{\cal
V}}V_{up}}{\partial^2_{{\cal V}}V_{0}}\right\rangle_0\,.
\end{equation}
Here it is explicit that the steeper the uplifting the larger the
shift is, and also the fact that it is inverse to the mass of the
volume modulus (non canonical normalized), given by the second
derivative of the potential.

\section{Explicit scenario}

Philosophically the proposed constrain (\ref{const}) seems to be
well motivated, but still we should do better in order to establish
how good it is a as discrimination tool for an effective uplifting
mechanism. Here we will do so using a single model of moduli
stabilization and a couple of potential uplifting mechanisms. The
first reason to chose this particular model is the fact that one of
the uplifting mechanisms to be shown have been studied previously
with the conclusion that it cannot lead to a Minkowski vacuum
\cite{mioLVS}. The second reason is that, so far, no explicit
numerical dS/Minkoski vacuum appear to be reported in the literature
and therefore we enjoy the opportunity to try to look for such using
a new uplifting mechanism proposed in ref.\cite{UpliftDilaton}.
\\
In general the LVS leading potential results from an expansion in
powers of the volume and exponentials from nonperturbative sources.
Then, with a K\"ahler potential with a leading dependency of the
form $K\sim-2 Log({\cal V})$, the potential scales with the volume
as
\begin{equation}\label{OriginalPotPara}
V_0\sim {\cal V}^{-3}\,.
\end{equation}
More precisely $\alpha^\prime$ corrections change the no-scale
nature of the model inducing a positive term that exactly go with
this power. The minimization process makes that all terms in the
leading potential at the end scale in the same way. As shown in the
previous section the behavior of the uplifting term is encoded in
the following potential
\begin{equation}
V_{up}= \nu {\cal V}^{-\gamma}\,,
\end{equation}
$\nu$ an amplitude modulated by tunable parameter or even VEV's of
other moduli. Then, eq.(\ref{shiftGen}) leads to the explicit
expression,
\begin{equation}\label{shiftExpl}
\frac{\delta {\cal V}}{{\cal V}} \approx \frac{\gamma}{12}\,,
\end{equation}
where we have used the requirement of a vanishing cosmological
constant, i.e., $|\langle V_{up}\rangle|\approx|\langle
V_{0}\rangle|$. After plugging this in the constrain (\ref{const}),
in an expansion in $\delta {\cal V}$ up to leading order, and
regarding again a vanishing cosmological constant we have the
following condition
\begin{equation}\label{Gbound}
\gamma^2\ll 12\,,
\end{equation}
stating a limit on the possible dependency of the uplifting on the
volume. Notice that this bound is stronger than the one that is
obtained by simply requiring that the shift, eq.(\ref{shiftExpl}),
be much more smaller than the volume it self.
\\
To check this result let us take the model proposed by Cremades et
al. in \cite{LVSMatter}, regarding three D7-branes, two of them on
top of each other, wrapped on a small 4-cycle in an orientifold
compactification. On the brane left alone a magnetic flux is turned
on generating a chiral spectrum with gauge group $SU(2)\times U(1)$,
the $U(1)$ being pseudo-anomalous. More precisely strings stretching
between the stack of branes and the single magnetized one generate
$\varphi$ fields transforming as $(\Box,1)$, while the ones
stretching between the orientifold image of the stack and the
magnetized brane, fields $\tilde\varphi$, transform as
$(\overline\Box,1)$. The K\"ahler modulus, $t$, parametrizing the
4-cycle where the branes wrap gets a nonlinear $U(1)$ charge due to
a Grenn-Schwarz mechanism for the cancelation of anomalies.
Practically one can work in a SUSY preserving mesonic direction,
canonically normalized $\phi=\sqrt{\varphi\tilde\varphi}$. Then
normalizing the charges such that $q_\phi=-1$ and $\delta^t=2/a$, a
gauge invariant nonpertubative ADS superpotential can be generated
\begin{equation}
W=W_{SC}+A\frac{e^{-a t}}{\phi^2}\,.
\end{equation}
The $W_{SC}$ part is a flux induced superpotential responsible for
the stabilization of the dilaton and complex structure moduli, at
SUSY preserving points, and the amplitude $A$ might also depend in
those fields. Regarding a Swiss-cheese like manifold with two
4-cycles: a large one, $T$, parametrizing the whole size of the
manifold and a small one, $t$, parametrizing the small cycle where
the branes are wrapped. The K\"ahler potential for the system is
given by
\begin{equation}
K=-2log({\cal V}+\hat \xi)+\frac{Z(t)}{{\cal V}^\eta}|\phi|^2\,,
\end{equation}
where ${\cal V}=T_r^{3/2}-t_r^{3/2}$, $Z$ is a function of the
complex structure and $t$, and $\hat \xi$ a $\alpha^\prime$
correction that depends on the dilaton modulus. The subindex $r$
denotes the real part but we omit hereafter such a notation as the
solutions turn out to be real for the values chosen in the numerics.
The gauge kinetic function for the $U(1)$ is given by $f_X=t$,
disregarding irrelevant but possible dependencies on the dilaton
induced by the magnetic flux. This leads to a D-term part of the
scalar potential of the form, at leading order in an expansion in
inverse powers of $T\approx {\cal V}^{2/3}$,
\begin{equation}
V_D\sim\frac{1}{t}\left(\frac{3}{2a}\frac{\sqrt{t}}{T^{3/2}}-\frac{Z(t)}{T^{3\eta/2}}
|\phi|^2\right)^2\,.
\end{equation}
The approximate cancelation of this potential fixes the value of
$\phi$ at $\langle |\phi|^2\rangle\approx \frac{3\sqrt{t_r}}{2a
Z(t_r)T_r^{3/2(1-\eta)}}$. Plugging this value in the F-term part of
the scalar potential, eq.(\ref{FtermPot}), leads to a potential for
the moduli with a LVS solution \cite{LVSMatter}
\begin{eqnarray}\label{ModPot}
V_{mod}&=&\frac{8 Z(t)^2 a^4 A^2 e^{-2a\,t}}{27 \sqrt{t}
T^{3\eta-3/2}}\cr &&-\frac{4\sqrt{t}Z(t)a^2 W_{SC} A e^{-a\,t}}{3
T^{3/2(1+\eta)}}+\frac{3 W_{CS}^2\hat \xi}{2T^{9/2}}\,,
\end{eqnarray}
regarding real values for all quantities, and whose VEV is negative.

\subsection{Uplifting proposal with extra matter}

Actually strings stretching between the magnetized brane and its
orientifold image generate fields $\rho$ singlets of $SU(2)$ and
doubly charged under $U(1)$, such that a coupling between this field
and $\phi$ is possible,
\begin{equation}
W_{up}=\frac12 m \rho \phi^2\,.
\end{equation}
Now an extra term in the potential appears from the $\rho$ F-term,
\begin{equation}
V_{up}\sim \frac{1}{4Z(t)T^{3(1-\eta/2)}}m^2 |\phi|^4\sim
\frac{m^2}{{\cal V}^{4-3\eta}}\,.
\end{equation}
Then by taking $m^2\sim {\cal V}^{1-3\eta}$ this term might cancel
the negative contribution from the moduli potential. However, a well
motivated value for the modular weight is $\eta=2/3$ that implies a
dependency of the uplifting in the volume with $\gamma=2$. Although
it seems plausible that such value still satisfy the condition
(\ref{Gbound}) by numerical computation, in a previous work
\cite{mioLVS}, the author checked that in fact such uplifting, with
this modular weight, is not able to generate Minkowski vacua.

\subsection{Dilaton dependent uplifting proposal}\label{Dilatondepsec}

An interesting possibility is one where the uplifting appears
depending on the dilaton modulus \cite{UpliftDilaton}. This field is
supposed to be already stabilized via fluxes at SUSY preserving
points, so such a term should come necessarily suppressed compared
to the flux induced potential \cite{mioLVS,mioLight} as can be
inferred also from the fact that it should be of the order of the
potential for the K\"ahler moduli in order to cancel the
cosmological constant. Such a term can be generated via gaugino
condensation in a D3-brane and/or E(-1)-instantons at a singularity
whose size is parametrized by a blow-up mode, $Q$. The presence of
magnetic fluxes generates a shift in in the gauge kinetic function,
given by the dilaton, and introduces a dependency on the blow-up
mode in an ADS superpotential
\begin{equation}\label{Wdilaton}
W_{up}=A_{up}e^{-a_{up}(S+h_{up}Q)}\,.
\end{equation}
The blow-up mode is stabilized at a nearly vanishing point by D-term
dynamics, something that we do not discuss in detail here but simply
introduce in the scalar potential as
\begin{equation}
V_{D,Q}=\frac{1}{8\pi(S_r+h_{up}Q_r)}\left( q_Q\frac{Q_r}{{\cal
V}}\right)^2\,,
\end{equation}
where we regard a K\'ahler potential for $Q$ of the form $(Q+\bar
Q)^2/{\cal V}$, $q_Q$ the corresponding $U(1)$ charge for $Q$. The
induced uplifting, coming from the $Q$ F-term, then is
\cite{UpliftDilaton}.
\begin{equation}\label{Suplift}
V_{up}\approx(a_{up}h_{up}A_{up})^2\frac{e^{-2a_{up}S_r}}{{\cal
V}}\,.
\end{equation}
Thus in this case $\gamma=1$ and the condition (\ref{Gbound}) is
clearly satisfied.

\section{Numerical results}
The aim now is to check whether this uplifting indeed does the job
or not. The study seems complicated due to the fact that the dilaton
appears as an important dynamical factor in the game. However,
regarding these moduli as stabilized at SUSY preserving point we can
rely on the results given by a model where the dilaton and complex
structure are just frozen \cite{mioLVS,mioLight}. Also in order to
clear up the ideas we suppose that there is no coupling with the
field $\rho$, i.e, $m=0$, for example due to some extra symmetry
forbidding it. In this case $\langle \rho\rangle=0$ with a non
tachyonic mass, so we do not report any result from this sector
being irrelevant for our study.
\\
Under this setup we set the following parameters and functions:
\begin{center}
\begin{tabular}{|c|c|c|c|}
  \hline
  $W_{CS}$ & $\hat\xi$ & $a$ & $Z(t)$ \\  \hline  \hline
  $10$  & $2$ & $3\pi$ &$\sqrt{t}$ \\
  \hline
\end{tabular}
\end{center}
For the dilaton we fix $S_o=1$ and an extra contribution in the
K\"ahler potential from the Dilaton and Complex Structure is fixed
to be $K_{SC}=-2Ln(2)$, whose value is actually irrelevant here but
is chosen having in mind a functional form like $K_{SC}=-Ln(S+\bar
S)-Ln(U+\bar U)$. The original vacuum is obtained by setting
$A_{up}=0$ in eq.(\ref{Wdilaton}) or simply ignoring the the $Q$
sector, finding real valued solutions for the VEV's shown in table
\ref{VEV0}.
\begin{table}[H]
\begin{center}
\begin{tabular}{|c|c|c|c|}
  \hline
  $T$ & $t$ & $\phi$ & $V_0$ \\  \hline  \hline
  $2.39\times 10^6$  & $3.26$ & $1.09\times 10^{-2}$ &$-9.44\times 10^{-29}$ \\
  \hline
\end{tabular}
\caption{VEV's at the original vacuum in natural units.}\label{VEV0}
\end{center}
\end{table}
\vspace{-0.3cm}
We report only the first two figure digits but the
existence and localization of the reported minima have been checked
up to the $30$th decimal place. The canonical normalized masses at
this point are given in table \ref{canMass0}.
\begin{table}[H]
\begin{center}
\begin{tabular}{|c|c|c|c|}
  \hline
  $m_{3/2}$&$m_\phi$ & $m_t$ & $m_T$  \\  \hline  \hline
  $3.25\times 10^{6}$&$1.44\times 10^{10}$  & $2.0\times 10^8$ & $85.7$  \\
  \hline
\end{tabular}
\caption{Canonical masses in $TeV$ units at the original vacuum.
$m_{3/2}$ the gravitino mass. For the $t$ field both, real and
imaginary, parts get roughly the same mass. The imaginary component
for $T$ is massless, expected to be stabilized by quantum
correction, and the imaginary part of $\phi$ is the would-be
Goldstone for the broken $U(1)$ symmetry.}\label{canMass0}
\end{center}
\end{table}
\vspace{-0.3cm}
Introducing the uplifting sector with parameters
\begin{center}
\begin{tabular}{|c|c|c|c|}
  \hline
  $A_{up}$ & $a_{up}$ & $h_{up}$ & $q_Q$ \\  \hline  \hline
  $-3.954\times 10^{-5}$  & $4\pi$ & $1/2$ &$1$ \\
  \hline
\end{tabular}
\end{center}
showing in $A_{up}$ up to the tuned decimal in order to get the
tuning for the cosmological constant below. The value for $A_{up}$
is sightly above to the one estimated by setting the term in
eq.(\ref{Suplift}) to cancel the original cosmological constant
mainly due to the shift in the position of the vacuum. We should
mention that although a positive value for $A_{up}$ seems to do also
the uplifting, in practice the phase introduced induces a tachyonic
mass for the $Q$ with real valued VEV's. Having said that we find
the new vacuum reported in table \ref{VEV}.
\begin{table}[H]
\begin{center}
\begin{tabular}{|c|c|c|c|}
  \hline
  $T$ & $t$ & $\phi$ &$Q$ \\  \hline  \hline
  $2.68\times 10^6$  & $3.28$ & $1.05\times 10^{-2}$&$6.27\times 10^{-7}$  \\
  \hline
\end{tabular}
\caption{VEV's at the uplifted vacuum in natural units.}\label{VEV}
\end{center}
\end{table}
The VEV for the potential is indeed uplifted to $3.42\times
10^{-32}$, in natural units, with a tuning at the level of almost
one part in ten thousand from a tuning in $A_{up}$ in one part in
five hundred. A better tuning would be justified by scanning the
string landscape of fluxes, as the amplitude $A_{up}$ in general
depend on the complex structure and dilaton moduli.
\\
The canonical masses are shown in table \ref{canMass}.
\begin{table}[H]
\begin{center}
\begin{tabular}{|c|c|c|c|}
  \hline
 $m_\phi$ &$m_Q$& $m_t$ & $m_T$  \\  \hline  \hline
  $1.32\times 10^{10}$  & $7.26\times 10^{9}$ & $1.7\times 10^{8}$&$54.4$  \\
  \hline
\end{tabular}
\caption{Canonical masses in $TeV$ units at the uplifted vacuum.
$m_{3/2}=2.75\times 10^6TeV$ the gravitino mass. For the $t$ field
both, real and imaginary, parts get roughly the same mass. The
imaginary component for $T$ is massless, expected to be stabilized
by quantum correction, and the imaginary part of $\phi$ and $Q$ are
the would-be Goldstone for the broken $U(1)$
symmetries.}\label{canMass}
\end{center}
\end{table}
\vspace{-0.3cm}
As advertised in the beginning the masses get
lowered compared to the ones at original vacuum. The $\phi$ field,
being stabilized by the D-term dynamics, has variation in the mass
mainly due to the shift in the scale of gauge symmetry breaking
dictated by the VEV of $\phi$. For the K\"ahler moduli instead,
getting the mass from the F-term part of the potential, this is due
to two effects: one is the larger volume that lowers all the energy
scales associated to the scalar potential, in particular the
gravitino mass now with value $m_{3/2}=2.75\times 10^6TeV$. More
precisely the overall factor $e^{K}\sim {\cal V}^{-2}$ makes that a
change $\Delta T=2.9\times 10^5$ would explain a decrement in the
masses of the K\"ahler moduli of order 30\%, as exactly happens in
for the small $t$ modulus. However, in the case of the large modulus
$T$ the variation is almost in 40\% which is explained, instead, by
the contributions from the uplifting that make flatter the potential
in this direction.

\section{Conclusions}

With the intention of having a guide for possible superstring vacua
realizing appealing phenomenological features we propose a
criterium, eq.(\ref{const}), the uplifting mechanism, if any, should
satisfy in order to effectively generate a tiny dS cosmological
constant. Given that the uplifting mechanisms in the market so far
have a clear dependency on the compact manifold volume, namely a
monomial with negative power, this bound can be explored with better
detail in the case of LVS where also the scalar potential isolates
the dependency in such direction, in the moduli space, in a rather
universal way. This leads to a bound for the power in the uplifting
term, eq.(\ref{Gbound}).
\\
An attentive reader might already notice that this last bound should
be taken with caution as the original LVS potential, see
eq.(\ref{ModPot}), does not actually scales like the third power of
the volume being in fact the combination of three terms scaling
differently all of them. Indeed this constraint actually gives the
less conservative possibility in the sense that the actual behavior
of the potential is the competition of three term and therefore is
sightly bellow the supposed third power that would lead to lower
bound. The bound in any case gives an indication of the order for
which one expects the uplifting term not to work. In particular
seems to exclude powers larger than two, and leave two at the edge.
This is actually the power found in the uplifting term proposed in
\cite{LVSMatter} which have been studied numerically in the past
with the outshot of the impossibility of finding Minkowski vacua.
\\
A novel uplifting mechanism was proposed in ref.\cite{UpliftDilaton}
where the power in the volume is one making it a potential option
for doing the job. As a check for the bound found analytically we
perform the numerical study of the uplifting finding for the first
time a dS vacuum in a setup with a charged small K\"ahler modulus.
\\
It is important, however, to leave clear that this results are far
from being conclusive. Indeed, although the bound clearly excludes
$\gamma=3$ it is not obvious that the $\gamma=2$ value, like in the
first case showed in the letter, does not work. More precisely for
values close to two we expect model dependent factors to be
important. This actually is checked by studying the special case of
uplifting with hypothetical modular weights zero and minus one,
leading to uplifting terms scaling like $1/{\cal V}^2$ and $1/{\cal
V}^3$ respectively. In the first case, contrary to the finding in
\cite{mioLVS}, we are able to find a vacuum, although the volume
modulus mass is almost 80\% less than the original one. The later
case, forbidden by the bound, indeed fails in getting a Minkowski
vacuum, with a negative cosmological constant independently of the
value for the amplitude.
\\
As stated in the beginning, and also shown in the explicit example,
the potential once is uplifted is flatter leading to a smaller mass
in the volume direction. This gives the indication that a more
precise, although possibly more complicated, way of discriminating
good or bad uplifting mechanism is through their side effects on the
masses (see for example \cite{Wrase1,Wrase2} for works in this
direction). We hope to come to this point in a later and more
extended study on the issue.




\nocite{*}
\bibliographystyle{elsarticle-num}
\biboptions{numbers,sort&compress}
\bibliography{biblio}
\end{document}